\documentstyle[12pt]{article}
\let\section=\subsection     \let\subsection=\subsubsection                
  
\begin{document}
\begin{center}
  {\large \bf SOME NEW ASPECTS OF BIMODAL FISSION}\\[2mm]
  {\large \bf IN $^{\bf258}$Fm ISOTOPE}\\[5mm]
  A.~STASZCZAK and Z.~\L{}OJEWSKI \\[5mm]
  {\small \it  Department of Theoretical Physics, M. Curie--Sk\l{}odowska 
  University \\
  pl. M. Curie--Sk\l{}odowskiej 1, 20--031 Lublin, Poland\\[8mm] }
\end{center}

\begin{abstract}\noindent
Using the multidimensional dynamic--programming method (MDPM) in
the four--dimensional deformation space
$\{\beta_{\lambda}\}$ with $\lambda$=2, 4, 35 and 6 we were able to study
evolution of the action integral of the fissioning nucleus $^{258}$Fm.
We found the second minimum on the cross--section of the action
integral for $\beta_2 \approx 1$ , what we interpret as the
dynamical evidence of the bimodal fission in this heavy Fm isotope.
\end{abstract}


\section{Introduction}

Experimentally known values of the spontaneous fission
half--lives $T_{sf}$ of nine even--even Fm isotopes (N = 142,
144, ..., 158) form approximately two sides of an acute--angled
triangle with a vertex in N = 152. The rapid changes in
$T_{sf}$ on both sides of $^{252}$Fm are particularly dramatic
for the heavier Fm isotopes, where $T_{sf}$ descends by about ten
orders of magnitude when one passes from $^{254}$Fm to $^{258}$Fm.

Nuclei in the vicinity of $^{258}$Fm reveal another two
peculiarities in spontaneous fission properties. The fission
process becomes symmetric with the narrow mass distributions
(see e.g. review \cite{H}) and the total--kinetic--energy (TKE)
distributions are not Gaussian, but instead are best described
as a sum of two Gaussians. For $^{258}$Fm the TKE distribution has
peaks at about 200 and 235 MeV. It was postulated \cite{HW} that for 
$^{258}$Fm the observed high--kinetic--energy peak corresponds to
fission through a scission configuration of two touching spherical
fragments, with the maximum of Coulomb repulsion. The
low--kinetic--energy peak was thought to correspond to fission through
a~conventional scission configuration of two elongated fragments. 

The investigations focused on a~multi--valley
structure of potential--energy surfaces for nuclei near Fm have
been undertaken by a~lot of authors. A~particularly detailed
studies were performed by a~Los Alamos group \cite{MN1,MN2,MN3} and a~Warsaw
group \cite{CR1,CR2}. Comparison of used models and results obtained by the 
both groups one may find in \cite{MN3} and in a review \cite{MN4}.

The descriptions of the bimodal fission in $^{258}$Fm proposed
by the Los Alamos and Warsaw groups are very similar. Roughly
speaking in both cases authors present two fission path on the 
potential--energy surfaces. One trajectory goes into the valley
which corresponds to more compact shapes (CS) of the nucleus and
the other to the valley which corresponds to more elongated
shapes (ES). However, the both presented descriptions of the bimodal 
fission have one common deficiency --- they are totally static, focused 
only on the multi--valley structure of potential--energy surfaces.

The main goal of this paper is to present a~dynamic analysis of
the bimodal fission in $^{258}$Fm. Thus, we focused our attention
on an action--integral of the fissioning nucleus in a
multi--dimensional deformation space.

The used model is described in sect. 2, the results and discussion
are given in sect. 3.

\section{Model}

To calculate the action--integral along a path $L(s)$ in the 
multi--dimensional deformation space $\{X_\lambda\}$ we
used the one--dimensional WKB semiclassical approximation
\begin{equation}
 S(L) = \int^{s_2}_{s_1} \left\{{2 \over \hbar^2} \, B_{\rm eff}(s) 
 [V(s) - E]\right\}^{1//2} ds\,,
\end{equation}
where an effective inertia associated with the fission motion
along the path $L(s)$ is
\begin{equation}
B_{\rm eff}(s) = \sum_{\lambda,\mu} \, B_{X_\lambda X_\mu} \,
  {dX_{\lambda} \over ds} {dX_\mu \over ds}\,.
\end{equation}

In above equations $ds$ defines the element of the path length in
the $\{X_\lambda\}$ space. The integration limits $s_1$ and
$s_2$ correspond to the entrance and exit points of the barrier
$V(s)$, determined by a~condition $V(s) = E$, where $E$ is the
penetration energy of the fissioning nucleus. 

The potential energy $V$ is calculated by the 
macroscopic--microscopic model
exactly the same like that used by the Warsaw group \cite{CR2}. For the
macroscopic part we used the Yukawa--plus--exponential
finite--range model \cite{KN} and for microscopic part the Strutinsky
shell correction, based on the Woods--Saxon single--particle
potential with ``universal'' variant of the parameters \cite{CD}. The
single--particle potential is extended to involve residual
pairing interaction, which is treated in the BCS approximation.
The inertia tensor $B_{X_\lambda X_\lambda}$, which
describes the inertia of the nucleus with respect to change of
its shape, is calculated in the cranking
approximation (cf. e.g. \cite{BP}). The penetration energy $E =
V(X^0_\lambda)$ + 0.5~MeV is defined as a sum of a~ground--state
energy $V(X^0_\lambda)$ at the equilibrium deformation $X^0_\lambda$
and a zero--point energy in the fission direction, equal 0.5 MeV.

Dynamic calculations of the spontaneous fission half--lives
$T_{sf}$ are understood as a quest for a~least--action
trajectory $L_{\rm min}$ which fulfills a principle of least--action
$\delta[S(L)]~=~0$. To minimize the action integral (1) we used the
multidimensional dynamic--programming method (MDPM).  Originally this
method was used only for two--dimensional deformation space \cite{BP}.
We extended the model up to four degrees of freedom.

The schematic Fig. 1 demonstrates how our model, the MDPM, works.
Since the macroscopic--microscopic method is not analytical, it is
necessary to calculate the potential energy and all components of 
the inertia tensor on a~grid in the multidimensional space spreaded by a set
of deformation parameters $\{X_\lambda\}$.
We select one coordinate $X_0$ from this set. This
coordinate (e.g. elongation parameter) is related in a linear
way to the fission process. In Fig. 1 for each point on a $X_0$
axis the rest of the coordinates $\{X_{\lambda-1}\}$
are represented for simplicity by a~``wall--plane'' of only two
coordinates $X_1$ and $X_2$.

To find the least--action trajectory $L_{\rm min}$ between the
turning point $s_1$ and $s_2$ we proceed as follows. First,
from the entrance point to the barrier $s_1$ we calculate the
action--integrals to all grid points in the nearest ``wall''
at $X_0 = 1$. In the next step we come to the ``wall'' at $X_0
= 2$ and from each grid point in this ``wall'' calculate the
action--integrals to all grid points in the ``wall'' at $X_0 =
1$ (see Fig. 1). The trajectories started from each grid point at
$X_0 = 2$ passing through all grid points in ``wall'' at $X_0 =
1$ and terminated in the point $s_1$ form a~bunch of paths. From
each such a~bunch we choose the path with minimal
action--integral and keep them in the memory. At the end of this
step we have the least--action integrals along trajectories
which connect the starting point $s_1$ with all grid points in
the ``wall'' at $X_0 = 2$. After that we repeat this procedure
for all grid points at $X_0 = 3$ and again we obtain all
least--action--integrals along trajectories starting from point
$s_1$ with ends at each grid point in the ``wall'' $X_0 = 3$. We
repeat it until we exceed the $n$--th ``wall''; a~last one before 
the exit point from the barrier $s_2$. After that we proceed to the
last step of our method. We calculate action--integrals between
all grid points in the last ``wall'' at $X_0 = n$ and the exit point
$s_2$; the minimal one among them corresponds to the searched
trajectory of the least--action--integral $L_{\rm min}$.

Our dynamical calculations are performed in the
four--dimensional deformation space spread by  $\beta$--shape
parameters appearing in an expansion of a nuclear radius in spherical
harmonics. So, we have  $\{X_\lambda\} \equiv \{\beta_\lambda\}$, 
with $\lambda =$ 2,4,35 and 6. Our deformation space is very similar to used 
in ref. \cite{CR2}. However, in opposite to ref. \cite{CR2}, we collect the
reflection--asymmetry parameters $\beta_3$ and $\beta_5$ in one parameter
$\beta_{35} \equiv (\beta_3, \beta_5{=}0.8\beta_3)$, according to the
average trajectory in a $(\beta_3, \beta_5)$ plane for nuclei in Fm region.

The potential energy $V$ and ten components of the symmetric
inertia tensor $B_{\beta_\lambda\beta_\lambda}$ are calculated
microscopically for
\begin{eqnarray}
&& \beta_2 = 0.15 (0.05) 1.30\,,\footnotemark \nonumber \\
&& \beta_4 = -0.08 (0.04) 0.36\,, \\
&& \beta_{35} = 0.00 (0.05) 0.25\,, \nonumber \\
&& \beta_6 = -0.12 (0.04) 0.12 \nonumber
\end{eqnarray}
\footnotetext {Here we use the notation $x=x_i(\Delta x)x_f$ which means
               that $x$ goes from $x_i$ through $x_f$ with step $\Delta x$.}
i.e. in the 24 $\times$ 12 $\times$ 6 $\times$ 7 = 12096 grid
points. In our calculation a~quadrupole deformation $\beta_2$
plays a~role of the coordinate $X_0$ from Fig. 1.

\section{Results and discussion}

Fig. 2 displays the cross--sections of the action--integral 
(in units of $\hbar$) of the nucleus $^{258}$Fm, obtained according
to our MDPM model for the different values of parameter $\beta_2$. 
To show a~three--dimensional structure of
these cross--sections we reduced the results to the 
two--dimensional $(\beta_4, \beta_{35})$ contour plots. This
we accomplish by plotting at each point a~minimal value of
action--integral with respect to a~third coordinate $\beta_6$.
Thus, in Fig. 2 one can see an evolution of the action--integral
in the fission process of $^{258}$Fm.

The two--dimensional contour diagrams show fairly smooth
surfaces with one minimum. On all contour diagrams a~position of
the minimum, in a~lower--left corner, is almost the same. 
If we join the minima from all contour maps, we obtain
a~dynamical fission trajectory in our four--dimensional
deformation space. The dynamical fission trajectory has
a~tendency to be close to a~straight line. It means a~reduction
of the effective inertia $B_{\rm eff}$ 
according to eq. (2) and leads to the smaller action--integral.
This dynamical fission trajectory goes to the
compact shapes (CS) (cf. e.g. Fig. 6 in ref. \cite{CR2}).

In the last contour map in Fig. 2 with the cross--section of the
action--integral for $\beta_2$ = 1.05 one can see in
an~upper--right corner a~second local minimum corresponding to
more elongated shapes (ES).
This second local minimum is more distinctly seen in Fig. 3
where we plot a contour diagram of the action--integral
obtained in the last step of the MDPM, when we
calculate the action--integral from all grid points on last
``wall'' to the exit point from the barrier $s_2$ (see Fig. 1). In
other words, in Fig. 3 we show a ``view'' of the last ``wall'' at
$\beta_2$ = 1.05 ``seen'' from the exit point $s_2$ with coordinates: 
($\beta_2$ = 1.10, $\beta_4$ = 0.05, $\beta_{35}$ = 0.01, 
$\beta_6$ = --0.06). 
The appearance of two minima on the cross--section of the action
integral, one global corresponding to more compact shapes (CS)
and another local corresponding to more elongated shapes (ES),
indicates that in the vicinity of $\beta_2$ = 1.05, before the
exit from the barrier, one can observe the begining of the two fission
modes in $^{258}$Fm .

Fig. 4 shows the potential energy $V$ calculated for 
$^{258}$Fm as a function of the deformation parameters $\beta_2$ and 
$\beta_4$. At each point $(\beta_2, \beta_4)$ we select the value of 
$V(\beta_2,\beta_4,\beta_{35},\beta_6)$ minimal with respect to
$\beta_{35}$ and $\beta_{6}$, where $\beta_{35}$ and $\beta_6$ are
taken only from the grid points, according to (3). On the right part of
the figure one can see two valleys. Entrances to the valleys of compact 
and elongated scission shapes are denoted by CS and by ES, respectively.

Fig. 4 also includes paths to fission. A nearly stright, solid line,
which starts in the first minimum (at $\beta_2\approx$ 0.25,
$\beta_4\approx$ 0.0) and goes to the CS valley, represents the
dynamical fission trajectory leading to compact shapes. From this
trajectory, close to the exit point from the barrier 
(at $\beta_2\approx$ 0.95, $\beta_4\approx$ 0.05), the fission
trajectory leading to the ES valley branches off (a dotted--dashed line).

Only for comparison we display in Fig. 4 a static fission trajectory
(i.e. the path of minimal potential energy for each $\beta_2$ value),
shown as a dashed line. This trajectory starts in the first minimum,
passes through the first saddle point  (at $\beta_2\approx$ 0.4,
$\beta_4\approx$ 0.1), then through the second minimum 
(at $\beta_2\approx$ 0.65, $\beta_4\approx$ 0.05), through the second
saddle point (at $\beta_2\approx$ 0.85, $\beta_4\approx$ 0.05), and
leads to the valley of elongated scission shapes. 

All fission barriers along the static and both dynamical trajectories
in Fig. 4 have two humps, in distinction to ref.\cite{CR2} where the static 
barrier has only one hump. This difference, despite the similar
macroscopic--microscopic models used, seems to come from minimization
procedure of the total energy of the nucleus in the multidimensional 
deformation space applayed by the Warsaw grup in \cite{CR2}. 

Our results of bimodal fission in $^{258}$Fm are in agreement with
those obtained by the Los Alamos group \cite{MN1,MN2,MN3,MN4}, where
the splitting of the trajectories appears also a little befor the exit
point from the barrier. This is in opposite to the Warsaw group
\cite{CR1,CR2} description where the fission trajectory splits behind the
exit point from the barrier.

\renewcommand{\topfraction}{0.0}
\renewcommand{\bottomfraction}{0.0}
\renewcommand{\floatpagefraction}{0.0}
\setcounter{topnumber}{20}
\setcounter{bottomnumber}{20}
\setcounter{totalnumber}{20}

\newpage
\centerline{\bf Figures Captions}

\begin{figure}[pht]
\caption[FIG1]{%
Schematic presentation of the MDPM method.
In the multidimensional deformation
space $\{X_\lambda\}$ we select the coordinate $\{X_0\}$
which is related in a linear way to the fission process. The points $s_1$
and $s_2$ correspond to entrance to the barrier and exit from the
barrier, respectively. See text for details.}
\label{FIG1}
\end{figure}

\begin{figure}[pht]
\caption[FIG2]{%
Evolution of the action--integral in the fission
process of $^{258}$Fm. Each contour map represents cross--sections 
of the action--integral, in $\hbar$ units, obtained for the different 
values of parameter $\beta_2$.}
\label{FIG2}
\end{figure}

\begin{figure}[pht]
\caption[FIG3]{%
View of the last contour map from Fig. 2, with 
cross--section of the action--integral at $\beta_2$ = 1.05, ``seen'' from 
the exit from the barrier point $s_2$. The letters CS indicate global  
minimum corresponding to more compact shapes, ES indicate second local 
minimum corresponging to more elongated shapes.}
\label{FIG3}
\end{figure}

\begin{figure}[pht]
\caption[FIG4]{%
Contour map of the potential energy of $^{258}$Fm, 
showing paths to fission. Solid line represents the dynamical fission
trajectory leading to compact shapes (CS), the dotted--dashed line denotes  
branches off second dynamical trajectory leading to elongated shapes (ES).
The dashed line represents the static fission trajectory.}
\label{FIG4}
\end{figure}

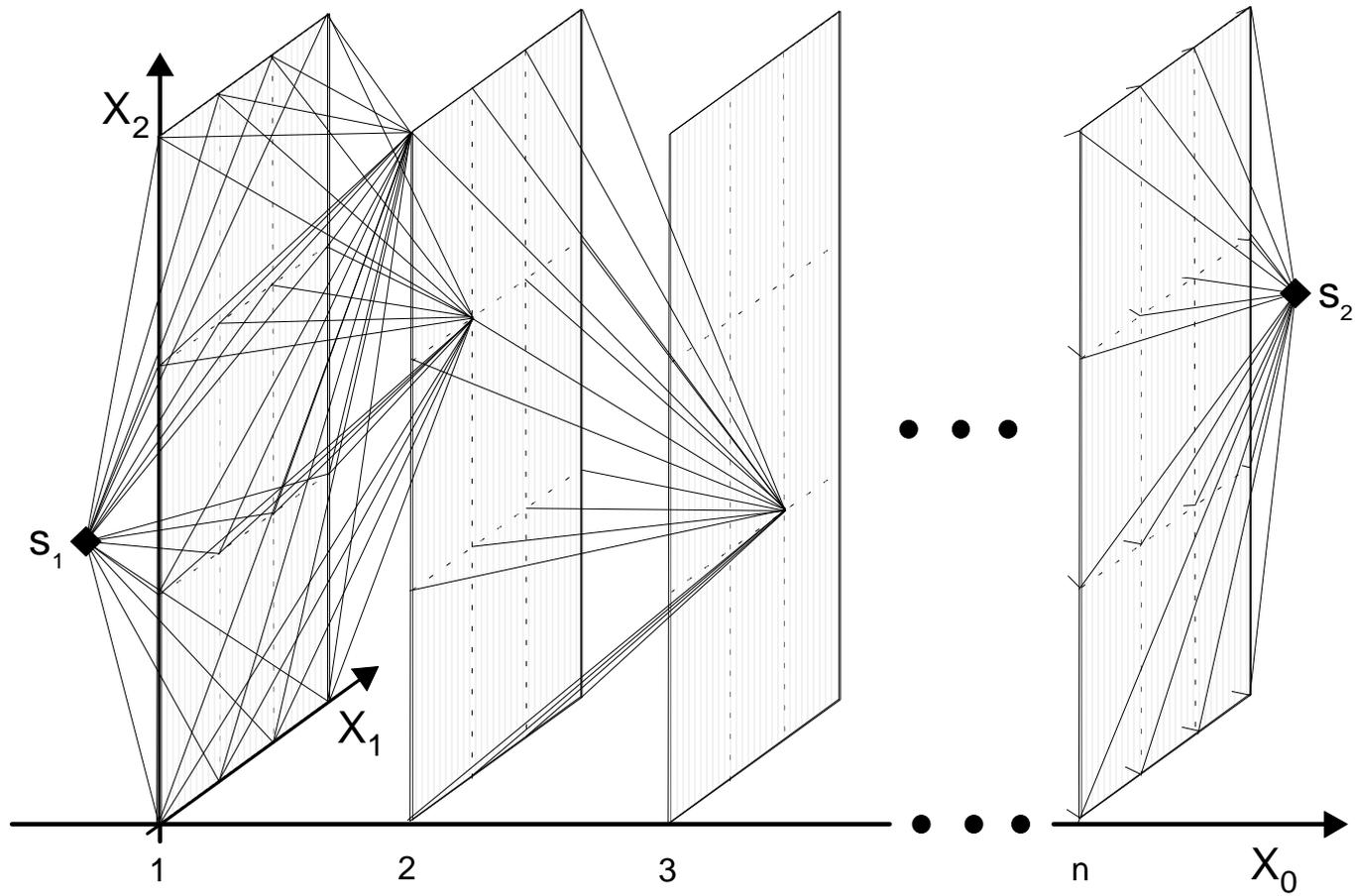
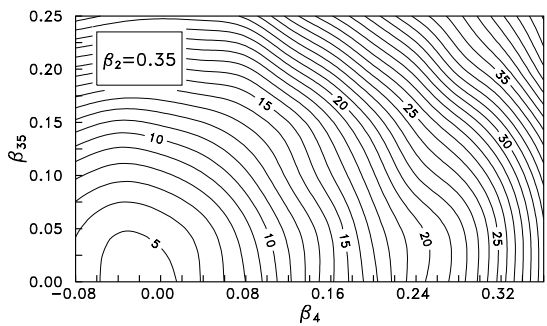
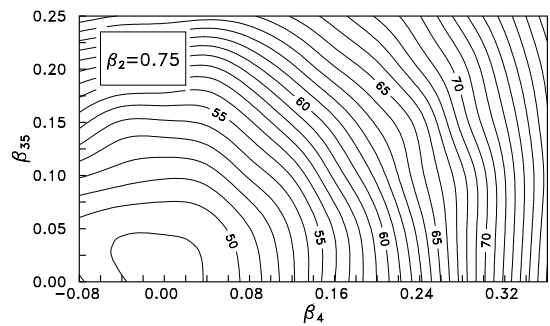
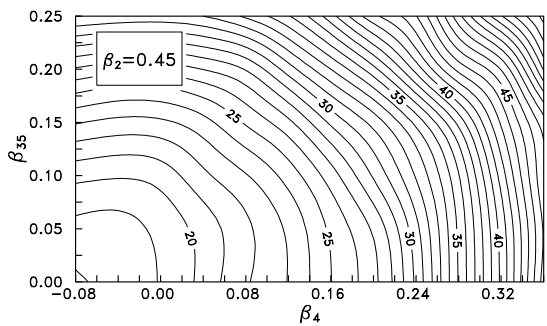
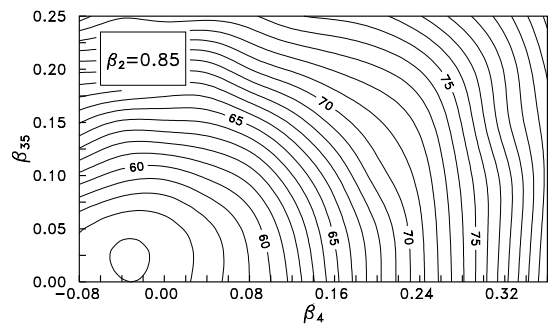
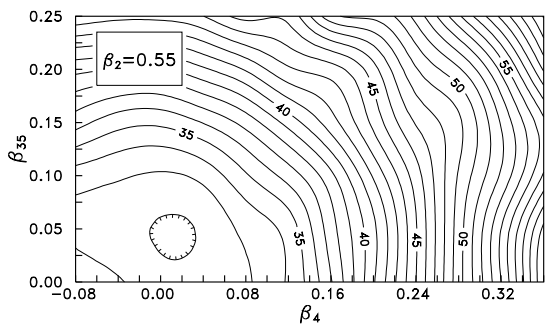
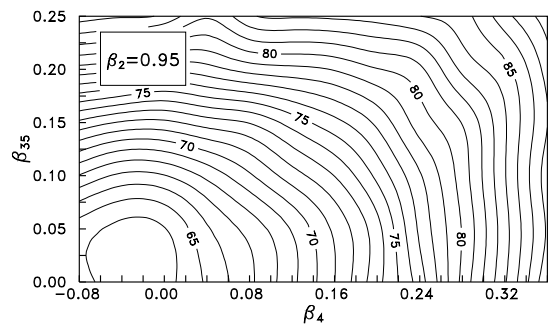
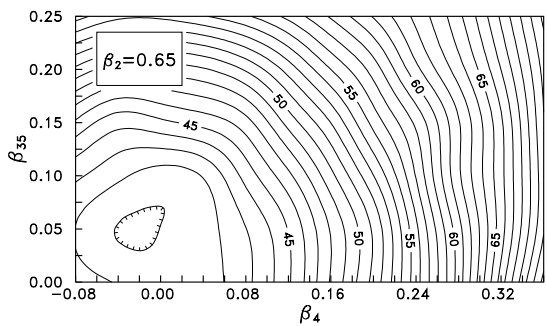
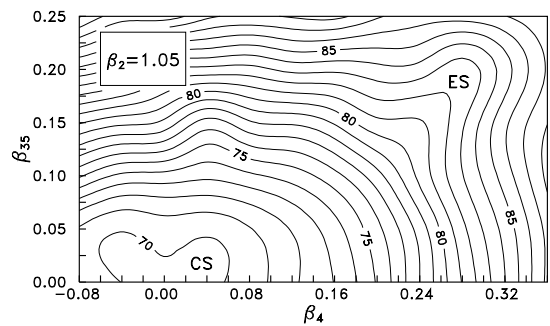
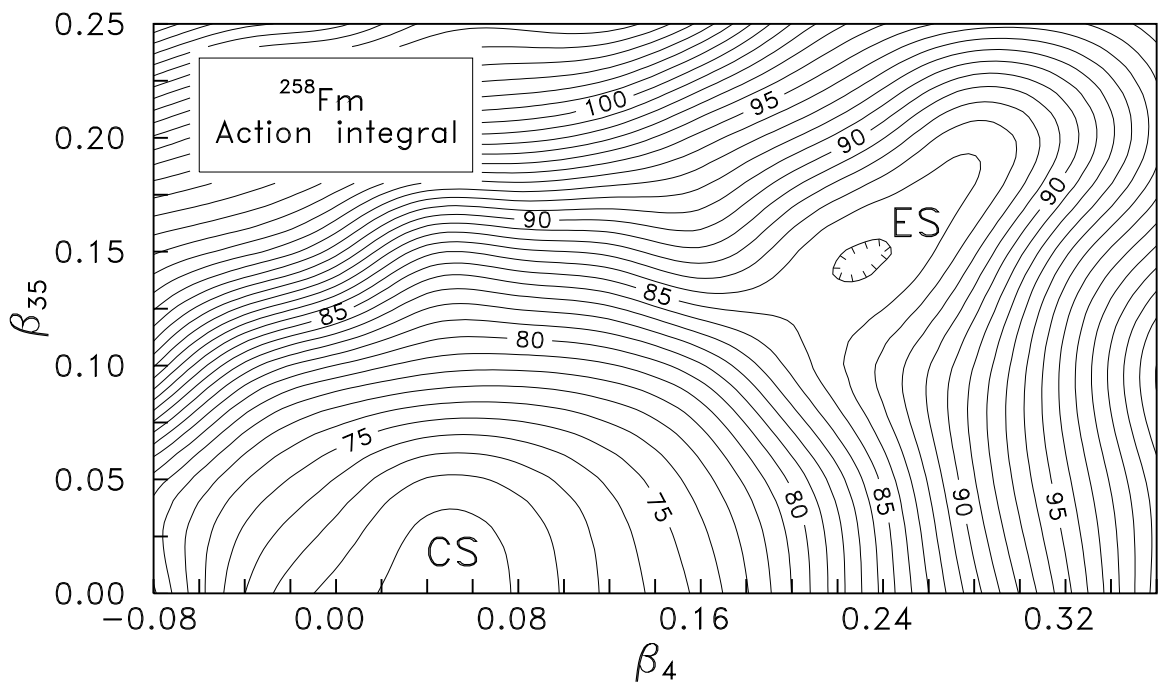
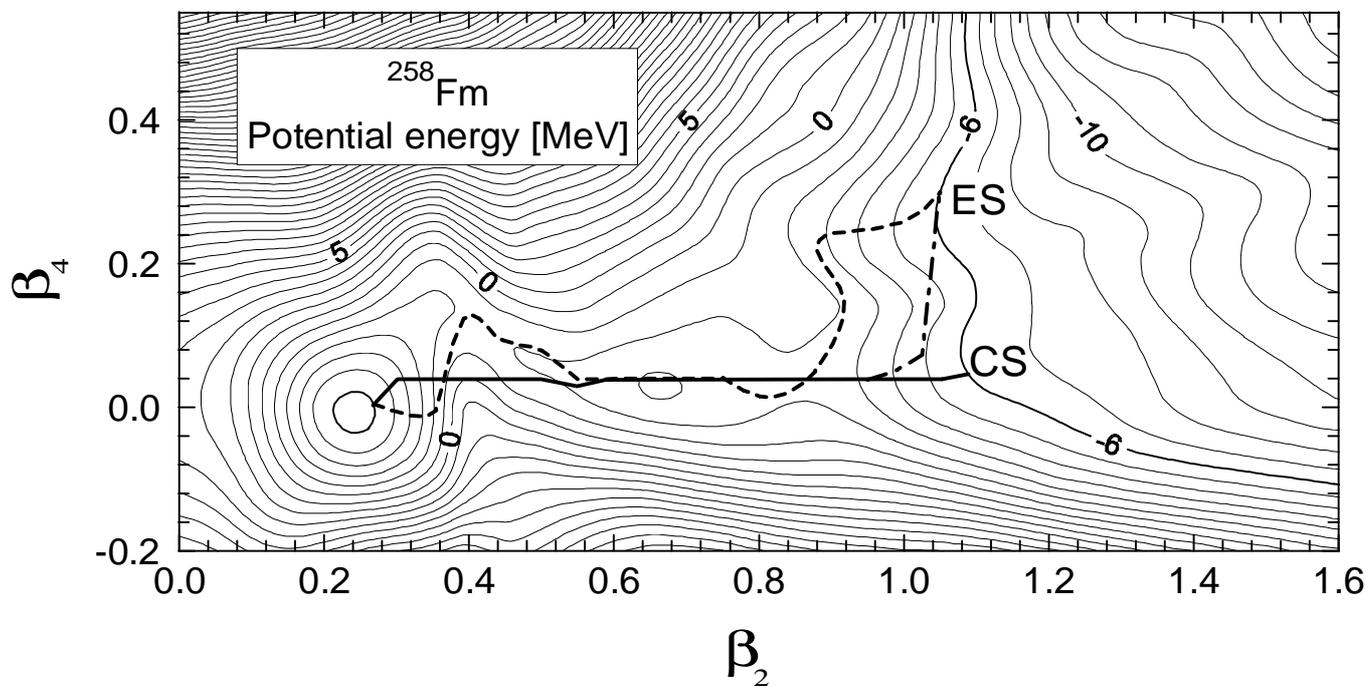
\end{document}